\documentclass[showpacs,twocolumn,eqsecnum,prb,amsmath,amssymb]{revtex4}

\usepackage{graphicx}
\usepackage{dcolumn}
\usepackage{bm}
\begin{document}
\title{
Electron-phonon interaction enhanced by antiferromagnetic and superconducting
fluctuations in cuprate oxide superconductors
}
\author{Fusayoshi J. Ohkawa}
\affiliation{Division of Physics, Graduate School of
Science,  Hokkaido University, Sapporo 060-0810, Japan}
\email{fohkawa@phys.sci.hokudai.ac.jp}
\date{\today}
%
\begin{abstract} 
An electron-phonon interaction arising from the modulation of the
superexchange interaction by phonons is studied  within the theoretical
framework of Kondo lattices. It is relevant in strongly correlated electron
liquids in cuprate oxide superconductors, which lie in the vicinity of the
Mott-Hubbard metal-insulator transition. It is
enhanced by antiferromagnetic and superconducting fluctuations, which are
developed mainly because of the superexchange interaction.  When the
enhancement of the electron-phonon interaction is large enough,  it  can
explain the softening of phonons and kinks in the quasiparticle dispersion in
cuprate oxide  superconductors. However, the superexchange interaction itself
must be mainly responsible for the formation of Cooper pairs.
\end{abstract}
\pacs{74.20.-z, 71.38.-k, 75.30.Et}
\maketitle

\section{Introduction}

It is an important issue to elucidate the mechanism of high-T$_c$
superconductivity occurring in cuprate oxides.\cite{bednorz} It was shown in
previous papers\cite{OhSC1,OhSC2} published in 1987 that the condensation of
$d\gamma$-wave Cooper pair  bound by the superexchange interaction can
explain observed $T_c$; $T_c$ for $d\gamma$ wave are definitely much higher
than those of other waves, as long as the on-site repulsion is so strong that
cuprate oxides with no dopings might be Mott-Hubbard insulators.   On the
other hand, two observations, the softening of
phonons\cite{McQ1,Pint1,McQ2,Pint2,Braden} and kinks in the quasiparticle
dispersion,\cite{Campuzano,lanzara,shen,Johnson,sato,timusk} imply the
relevance of an electron-phonon interaction.  In particular, the softening of
phonons is evidence that an electron-phonon interaction is strong in cuprate
oxides. One may argue that it must  be responsible for  high-T$_c$
superconductivity or, at least, it must play some role in the realization of 
high-T$_c$ superconductivity.\cite{lanzara,shen} 

Various experiments\cite{yasuoka,Ding,Shen2,Shen3,Ino,Renner,Ido1,Ido2} imply
or show the opening of anisotropic pseudogaps at temperatures above $T_c$ in
quasiparticle spectra of the so called optimal and under-doped cuprate oxide
superconductors. One may argue that kinks are caused by or are closely
related with what cause pseudogaps, rather than phonons. Antiferromagnetic
(AF) and superconducting (SC) fluctuations are developed in cuprate oxide
superconductors; not only antiferromagnetism and superconductivity themselves
but also the development of their fluctuations are  mainly  caused by the
superexchange interaction. It was shown in a previous paper\cite{OhPseudo}
that large  life-time widths, which are
mainly caused by well developed $d\gamma$-wave SC fluctuations so that they
are anisotropic or wave-number dependent, are responsible for anisotropic
pseudogaps; spectral weights around the chemical potential are swept away
because of large life-time widths.   Small kinks appear in calculated
spectra,\cite{OhPseudo} but they are too small to explain observed kinks. It
is difficult to explain kinks by AF or SC fluctuations. It is reasonable
that not only the softening of phonons but also kinks are caused by
an electron-phonon interaction. 

Doped {\em holes} mainly go into O ions. This implies that the
local charge susceptibility of $3d$ electrons on Cu ions is much smaller than
that of $2p$ electrons on O ions and charge fluctuation of
$3d$ electrons can never be developed. It is quite unlikely that the
conventional electron-phonon interaction, which directly couples with charge
fluctuations, plays a crucial role in cuprate oxide superconductors.  

A  necessary condition for a relevant
electron-phonon interaction is that it can work even in strongly correlated
electron liquids in the vicinity of the Mott-Hubbard 
metal-insulator transition or even when charge fluctuations are significantly
suppressed.  One of the most plausible ones is an electron-phonon interaction
arising from the modulation of the superexchange interaction by  phonons. 
It was pointed out that it
plays a role in phonon-assisted multi-magnon optical
absorption.\cite{lorenzana} 
It can couple directly with AF and SC fluctuations, so that it can be
substantially enhanced by AF and SC fluctuations.    One of the
purposes of this paper is to show that the electron-phonon interaction 
arising from the modulation of the superexchange interaction by  phonons is
relevant, at least, in cuprate oxide superconductors  where AF and SC
fluctuations are substantially developed.

\section{Formulation}
\subsection{Electron-phonon interaction}

It was shown in a previous paper \cite{oh-slave}
that  Gutzwiller's quasiparticle band \cite{gutzwiller} lies between the
lower and upper Hubbard bands \cite{hubbard} in metallic phases in the
vicinity of the Mott-Hubbard transition. 
Gutzwiller's quasiparticles are responsible for metallic properties. The
superexchange interaction arises from the virtual exchange of pair
excitations of electrons across the lower and upper Hubbard bands. As long as
the Hubbard splitting is significant, therefore, it works between Gutzwiller's
quasiparticles.

When we follow  previous papers\cite{OhSupJ0,OhSupJ1,OhSupJ2} and  we ignore
nonzero bandwidths of the lower and upper Hubbard bands, it is
straightforward to show that the virtual exchange process gives the following
exchange constant between  nearest-neighbor $i$th and $j$th Cu ions:
\begin{eqnarray}\label{EqSuperJ0}
J_{ij}  &=& - 4V_{i,[ij]}^2 V_{j,[ij]}^2
\left\{
\frac{2}{\epsilon_{dj}-\epsilon_{di}} \right.
\nonumber \\
&& \quad \times 
\left[
\frac{1}{(\epsilon_{di}+U  - \epsilon_{p[ij]})^2} 
- \frac{1}{(\epsilon_{dj}+U  - \epsilon_{p[ij]})^2}
\right] 
\nonumber \\ && + 
\frac{2}{(\epsilon_{di}+U  - \epsilon_{p[ij]})^2}
\frac{1}{\epsilon_{di}+U - \epsilon_{dj}}
\nonumber \\ && \quad  \left.
+ \frac{2}{(\epsilon_{dj}+U  - \epsilon_{p[ij]})^2}
\frac{1}{\epsilon_{dj}+U - \epsilon_{di}} \right\} ,
\end{eqnarray}
where the $3d$ levels of the $i$th and $j$th Cu ions are denoted by
$\epsilon_{di}$ and $\epsilon_{dj}$, the $2p$ level of the $[ij]$th O ion
between the two Cu ions by $\epsilon_{p[ij]}$, and the
hybridization energies between the Cu ions and the O ion by $V_{i,[ij]}$
and $V_{j,[ij]}$, respectively.  
When we put
$\epsilon_{di}\rightarrow\epsilon_{d}$,  
$\epsilon_{dj}\rightarrow\epsilon_{d}$,
and $V_{i,[ij]}=V_{j,[ij]}=V$, we obtain a wellknown one:
\begin{equation}\label{EqSuperJ}
J =-  \frac{4V^4}{(\epsilon_d+U-\epsilon_p)^2}
\left[\frac{1}{\epsilon_d+U-\epsilon_p}+\frac{1}{U} \right] .
\end{equation}
The variation of  $J_{ij}$ is given by 
\begin{eqnarray}
\Delta J_{ij} &=&
\frac{2V^4}{(\epsilon_d+U-\epsilon_p)^3}
\left[\frac{3}{\epsilon_d+U-v_p}+\frac{2}{U} \right]
\nonumber\\  && \hspace*{1cm} \times (\Delta \epsilon_{di}+\Delta
\epsilon_{dj}-2\Delta \epsilon_{p[ij]})
\nonumber \\ 
&& +
2\frac{J}{V} (\Delta V_{i,[ij]}+ \Delta V_{j,[ij]}) .
\end{eqnarray}
 When we take the $x$- and $y$-axes
along Cu-O-Cu bonds, 
variations of 
$\epsilon_{di}$, $\epsilon_{p[ij]}$ and $V_{i,[ij]}$ are given by
\begin{equation} \label{EqVar}
\Delta \epsilon_{di} \!=\! A_d \!\left[
{\bf e}_{x}\!\cdot\!({\bf u}_{i,x_+} \!\!-\! {\bf u}_{i,x_-})
\!+\! {\bf e}_{y}\!\cdot\! ({\bf u}_{i,y_+} \!\!-\! {\bf u}_{i,y_-})
\right],
\end{equation}
\vspace{-0.5cm} 
\begin{equation} 
\Delta \epsilon_{p[ij]} = A_p \left[
{\bf e}_{ij}\cdot({\bf u}_{i}-{\bf u}_{j}) \right],
\end{equation} 
\vspace{-0.5cm} 
\begin{equation} 
\Delta V_{i,[ij]}+ \Delta V_{j,[ij]}
= A_V \left[
{\bf e}_{ij}\cdot({\bf u}_{i}-{\bf u}_{j}) \right] ,
\end{equation} 
to linear order in displacement of ions, 
with $A_d$, $A_p$ and $A_V$ being constants, 
${\bf u}_i$  the displacement of the $i$th Cu ion, 
${\bf u}_{i,\xi_s}$ that of an O ion on the adjacent $s=+$ or $s=-$
side along the $\xi$-axis of the $i$th Cu ion,  
${\bf e}_x =(1,0)$, ${\bf e}_y =(0,1)$,
and
${\bf e}_{ij}=({\bf R}_i -{\bf R}_j)/|{\bf R}_i -{\bf R}_j|$, 
with ${\bf R}_i$ the position of the $i$th Cu ion.

Displacements of the $i$th Cu and the $[ij]$th O ions are given by
\begin{equation}\label{EqDispCu}
{\bf u}_i =
 \sum_{\lambda{\bf q}}
 \frac{\hbar v_{d,\lambda{\bf q}} } 
{\sqrt{ 2NM_d \omega_{\lambda{\bf q}}} } 
e^{i{\bf q}\cdot{\bf R}_i}
{\bm \epsilon}_{\lambda{\bf q}} \left(
b_{\lambda-{\bf q}}^\dag \!+\! b_{\lambda{\bf q}} \right), 
\end{equation}
\vspace{-0.3cm} 
\begin{equation}\label{EqDispO}
{\bf u}_{[ij]} =
\sum_{\lambda{\bf q}}
\frac{\hbar v_{p,\lambda{\bf q}}}
{\sqrt{2N M_p \omega_{\lambda{\bf q}}} } 
 e^{i{\bf q}\cdot {\bf R}_{[ij]} }
{\bm \epsilon}_{\lambda{\bf q}} \! \left(
b_{\lambda-{\bf q}}^\dag \!\!+\! b_{\lambda{\bf q}} \right), 
\end{equation}
with  ${\bf R}_{[ij]} \!=\! (1/2)({\bf R}_i \!+\! {\bf R}_j)$, 
$M_d$ the mass of Cu ions, $M_p$ the mass of O ions, $b_{\lambda{\bf q}}$ and 
$b_{\lambda-{\bf q}}^\dag$  annihilation and creation operators of phonons with 
polarization $\lambda$ and wave vector ${\bf q}$, $\omega_{\lambda{\bf q}}$ 
energies of phonons,  ${\bm \epsilon}_{\lambda{\bf q}}$ unit polarization
vectors, and $N$ the number of unit cells. The ${\bf q}$ dependence of
$v_{d,\lambda{\bf q}}$ and $v_{p,\lambda{\bf q}}$ can play a crucial role.
For example, 
$v_{d,\lambda{\bf q}}=0$ and 
$v_{p,\lambda{\bf q}} = O(1)$   
for breathing modes that bring no changes in
adjacent Cu-Cu distances.

The electronic part can be  well described by
the $t$-$J$ model on a square lattice:
\begin{eqnarray}\label{EqtJ}
{\cal H} &=& 
\epsilon_d \sum_{i\sigma}  d_{i\sigma}^\dag d_{i\sigma} 
- \sum_{ij\sigma} t_{ij} d_{i\sigma}^\dag d_{j\sigma} 
\nonumber \\ &&
-  \frac1{2}J 
\sum_{\left<ij\right>} ({\bf S}_i \cdot {\bf S}_j)   
+  U_{\infty} \sum_{i} n_{i\uparrow}n_{i\downarrow} ,
\end{eqnarray}
with the summation over $\left<ij\right>$ restricted to nearest
neighbors,
\begin{equation}
{\bf S}_i = \frac1{2} \sum_{\alpha\beta}  \left(
\sigma_x^{\alpha\beta} \!, \sigma_y^{\alpha\beta} \!,
\sigma_z^{\alpha\beta} \right) d_{i\alpha}^\dagger d_{i\beta},
\end{equation}
with $\sigma_x$, $\sigma_y$ and $\sigma_z$ the Pauli matrices, 
and $n_{i\sigma}= d_{i\sigma}^\dag d_{i\sigma}$. 
An infinitely large on-site repulsion, 
$U_\infty/|t_{\left<ij\right>}|\rightarrow +\infty$, 
is introduced to exclude any doubly occupied sites. 

According to Eq.~(\ref{EqVar}), there are two types of electron-phonon
interactions. Define an operator by
\begin{equation}\label{EqTwoSpin}
\!{\cal P}_\Gamma({\bf q}) = \frac1{2} \!
\sum_{{\bf q}^\prime}\eta_{\Gamma}({\bf q}^\prime) \! \left[
 {\bf S}\left({\bf q}^\prime \!\!+\! \mbox{$\frac{1}{2}$}{\bf q}\right)
\!\cdot {\bf S}\left(-{\bf q}^\prime \!\!+\! \mbox{$\frac{1}{2}$}{\bf q}
\right)\right] ,
\end{equation}
with
\begin{equation}
{\bf S}({\bf q}) = \frac1{\sqrt{N}} \sum_{\bf k\alpha\beta} 
\frac1{2}{\bm \sigma}^{\alpha\beta} d_{({\bf k} +\frac{1}{2}{\bf q})
\alpha}^\dag   d_{({\bf k} -\frac{1}{2}{\bf q}) \beta} ,
\end{equation}
with $ {\bm \sigma} = \left(\sigma_x , \sigma_y ,\sigma_z \right)$.
They are given by
\begin{eqnarray}\label{EqElPhP}
{\cal H}_p &=&
i C_p \sum_{\bf q} 
\frac{\hbar v_{p,\lambda{\bf q}}}
{\sqrt{2 N M_p \omega_{\lambda{\bf q}}}} 
\left(b_{\lambda-{\bf q}}^\dag + b_{\lambda{\bf q}} \right)
\nonumber \\ &&  \quad \times
\bar{\eta}_{s}({\bf q}) \sum_{\Gamma=s,d} 
\eta_{\Gamma}(\mbox{$\frac{1}{2}{\bf q}$}) 
{\cal P}_\Gamma({\bf q}), 
\end{eqnarray}
\vspace{-0.5cm} 
\begin{eqnarray} \label{EqElPhD}
{\cal H}_d &=&
i C_d \sum_{\bf q} 
\frac{\hbar v_{d,\lambda{\bf q}}}{\sqrt{2 N M_d \omega_{\lambda{\bf q}}}} 
\left(b_{\lambda-{\bf q}}^\dag + b_{\lambda{\bf q}} \right)
\nonumber \\ &&  \quad \times
\sum_{\Gamma=s,d} 
\bar{\eta}_{\Gamma}({\bf q})
{\cal P}_\Gamma({\bf q}) , 
\end{eqnarray}
with
\begin{equation}\label{EqCp}
C_p =  \frac{8 A_d V^4}{(\epsilon_d+U-\epsilon_p)^3}
\left[\frac{3}{\epsilon_d+U-\epsilon_p}+\frac{2}{U} \right], 
\end{equation}
\vspace{-0.3cm} 
\begin{equation} \label{EqCd}
C_d = - \frac{4A_pV^4}{(\epsilon_d \!+\! U \!-\! \epsilon_p)^3}
\left[\frac{3}{\epsilon_d \!+\! U \!-\! \epsilon_p} \!+\! \frac{2}{U} \right]
\!+\! \frac{ 2 A_V J}{V},  
\end{equation}
\begin{equation}
\bar{\eta}_{s}({\bf q}) = 
2\left[\frac{q_x}{q} \sin\left(\frac{q_x a}{2}\right) 
\!+\! \frac{q_y}{q} \sin\left(\frac{q_y a}{2}\right)\right] ,
\end{equation}
\vspace{-0.3cm} 
\begin{equation}
\bar{\eta}_{d}({\bf q}) = 
2\left[\frac{q_x}{q} \sin\left(\frac{q_x a}{2}\right) 
- \frac{q_y}{q} \sin\left(\frac{q_y a}{2}\right)\right] , 
\end{equation}
\begin{equation}  
\eta_{s}({\bf k}) = \cos(k_xa) + \cos(k_ya), 
\end{equation}
\vspace{-0.3cm} 
\begin{equation}
\eta_{d}({\bf k}) = \cos(k_xa) - \cos(k_ya),
\end{equation}
with $a$ the lattice constant. Here, we consider only longitudinal phonons or
${\bm \epsilon}_{{\lambda}{\bf q}} = (q_x,q_y,q_z) /q $ is assumed.


\subsection{Theory of Kondo lattices}

We follow the previous paper \cite{OhPseudo} to treat the infinitely large
$U_\infty$, 
 where a theory of Kondo lattice is developed.
A renormalized single-site approximation (SSA), which includes not only all the
single-site terms but also the Fock term $\Delta\Sigma({\bf k})$ due to the
superexchange interaction, is reduced to solving the Anderson model with the
infinitely large on-site repulsion $U_\infty$.
The self-energy of the Anderson model is expanded as
\begin{eqnarray}
\tilde{\Sigma}_\sigma(i\varepsilon_n) &=&
\tilde{\Sigma}(0) + (1-\tilde{\phi}_\gamma) i\varepsilon_n
\nonumber \\ && \quad 
+\sum_{\sigma^\prime}(1-\tilde{\phi}_{\sigma\sigma^\prime}) 
\Delta\mu_{\sigma^\prime} + \cdots,
\end{eqnarray}
with $\Delta\mu_{\sigma}$
infinitesimally small spin-dependent chemical potential shifts. Note that 
$\tilde{\phi}_\gamma=\tilde{\phi}_{\sigma\sigma}$.
The Wilson ratio is defined by
$\tilde{W}_s = \tilde{\phi}_s/\tilde{\phi}_\gamma$,
with $\tilde{\phi}_s=
\tilde{\phi}_{\sigma\sigma}-\tilde{\phi}_{\sigma-\sigma}$.
For almost half filling, charge fluctuations are suppressed so that
$\tilde{\phi}_c=
\tilde{\phi}_{\sigma\sigma}+\tilde{\phi}_{\sigma-\sigma} \ll 1$.
For such filling, $\tilde{\phi}_\gamma \gg 1$ so that
$\tilde{\phi}_s \simeq 2\tilde{\phi}_\gamma$ or $\tilde{W}_s \simeq 2$. The
dispersion relation of quasiparticles is given by 
\begin{equation}
\xi({\bf k}) = 
\frac{1}{\tilde{\phi}_\gamma}
\!\biggl[\epsilon_d \!-\! \sum_{j}t_{ij} e^{i{\bf k}\cdot
\left({\bf R}_i - {\bf R}_j\right) }
\!+\! \tilde{\Sigma}(0) \!+\! \Delta\Sigma({\bf k}) \!-\! \mu \biggr],
\end{equation}
with  $\mu$ the chemical potential.

The spin susceptibility is given by
\begin{equation}
\chi_s(i\omega_l,{\bf q}) =
\frac{2 \pi_s(i\omega_l,{\bf q})}{
1 \!-\! \left[ \frac1{2} J({\bf q}) \!+\! 
U_\infty \right]\pi_s(i\omega_l,{\bf q}) },
\end{equation} 
with $\pi_s(i\omega_l,{\bf q})$ 
the irreducible polarization function in spin
channels and 
\begin{equation}
J({\bf q})=2J \eta_s({\bf q}). 
\end{equation} 
The function $\pi_s(i\omega_l,{\bf q})$ is
divided into single-site  
$\tilde{\pi}_s(i\omega_l)$  and multi-site 
$\Delta\pi_s(i\omega_l, {\bf q})$ in such a way that
$\pi_s(i\omega_l,{\bf q}) = \tilde{\pi}_s(i\omega_l) + 
\Delta\pi_s(i\omega_l, {\bf q})$. 
In Kondo lattices, local spin fluctuations at different sites interact with
each other by an exchange interaction.  Following this physical picture, we
define an exchange interaction
$I_s(i\omega_l, {\bf q})$ by 
\begin{equation}\label{EqKondoSus}
\chi_s(i\omega_l, {\bf q}) =
\frac{\tilde{\chi}_s(i\omega_l)} 
{1 - \mbox{$\frac{1}{4}$}I_s(i\omega_l, {\bf q})
\tilde{\chi}_s(i\omega_l)} ,
\end{equation}
with 
\begin{equation}
\tilde{\chi}_s(i\omega_l) = \frac{2 \tilde{\pi}_s(i\omega_l)}
{1 - U_\infty \tilde{\pi}_s(i\omega_l) }
\end{equation}
the susceptibility for the mapped Anderson model. 
Then, we obtain 
\begin{equation}\label{EqIs}
I_s (i\omega_l, {\bf q}) = J({\bf q}) +  
2 U_\infty^2 \Delta\pi_s(i\omega_l, {\bf q}) .
\end{equation}
 The main part of  $2 U_\infty^2 
\Delta\pi_s(i\omega_l, {\bf q}) $ is an
exchange interaction arising from the virtual exchange of pair excitations of
quasiparticles.

When the Ward relation \cite{ward} is made use of,
the irreducible single-site three-point vertex function in spin channels, 
$\tilde{\lambda}_s(i\varepsilon_n,i\varepsilon_n \!+\! i\omega_l;i\omega_l)$,
is given by
\begin{equation}\label{EqThreeL}
 U_\infty 
\tilde{\lambda}_s(i\varepsilon_n,i\varepsilon_n+i\omega_l;i\omega_l)
= 2\tilde{\phi}_s / \tilde{\chi}_s(i\omega_l)  ,
\end{equation}
for $|\varepsilon_n| \rightarrow +0$ and 
$|\omega_l| \rightarrow +0$.
 We approximately use Eq.~(\ref{EqThreeL}) for 
$|\varepsilon_n| \alt k_BT_K$ and
$|\omega_l| \alt k_BT_K$, with $T_K$ the Kondo temperature defined by
\begin{equation} 
k_BT_K = \left[1/\tilde{\chi}_s(0)\right]_{T\rightarrow 0}.  
\end{equation}
The so called spin-fluctuation mediated interaction, whose single-site term
should be subtracted because it is considered in SSA, is given by 
\begin{equation}\label{EqIs*1}
\frac{1}{4} \left[
2\tilde{\phi}_s / \tilde{\chi}_s(i\omega_l)\right]^2
F(i\omega_l, {\bf q})   =
\tilde{\phi}_s^2 \frac{1}{4} I_s^* (i\omega_l, {\bf q}), 
\end{equation}
with 
\begin{equation}
F(i\omega_l, {\bf q}) = \chi_s(i\omega_l, {\bf q}) - 
\tilde{\chi}_s(i\omega_l),
\end{equation}
and
\begin{equation}\label{EqIs*2}
\frac{1}{4} I_s^*(i\omega_l, {\bf q}) =
\frac{ \frac{1}{4}
I_s (i\omega_l, {\bf q}) }
{1 - \frac{1}{4}I_s(i\omega_l, {\bf q})
\tilde{\chi}_s(i\omega_l) } .
\end{equation}
 Because of these equations, we call $I_s(i\omega_l, {\bf q})$ a 
{\em bare} exchange interaction, $I_s^*(i\omega_l, {\bf q})$ an
 enhanced one, and $\tilde{\phi}_s$ an effective three-point
vertex function in spin channels.
Intersite effects can be perturbatively considered in terms of
$F(i\omega_l, {\bf q}) $,
$I_s (i\omega_l, {\bf q})$ or 
$I_s^*(i\omega_l, {\bf q}) $
depending on each situation.


The enhanced one is  expanded as
\begin{equation}
I_s^*(i\omega_l,{\bf q}) = I_0^* + 2 I_1^* \eta_s({\bf q}) 
+ 2 I_2^* \eta_{s2}({\bf q}) + \cdots,
\end{equation} 
with 
\begin{equation}
\eta_{s2}({\bf q})= \cos\left[\left(k_x \!+\! k_y\right)a\right]
+\cos\left[\left(k_x \!-\! k_y\right)a\right]. 
\end{equation}
The nearest-neighbor $I_1^*$ is mainly responsible for the development of
 SC and charge bond-order (CBO) fluctuations.\cite{ComCBO} 
Because contributions from $|\omega_l|\alt k_BT_K$ are the most effective,
we ignore its energy dependence.
An effective SC  susceptibility is calculated in the
ladder approximation:
\begin{equation}\label{EqSCsus}
\chi_{\Gamma=d}^{\mbox{\tiny (SC)}}(i\omega_l,{\bf q}) =
\frac{ \pi_{d}^{\mbox{\tiny (SC)}}(i\omega_l,{\bf q})}
{1 + \frac{3}{4} I_1^* \tilde{W}_s^2
\pi_{d}^{\mbox{\tiny (SC)}}(i\omega_l,{\bf q})},
\end{equation}
for $\Gamma=d$ wave or $d\gamma$ wave, with 
\begin{eqnarray}
\pi^{\mbox{\tiny (SC)}}_{\Gamma}(i\omega_l,{\bf q}) &=&
\frac{k_B T}{N}\sum_{n{\bf k}} 
\eta_{\Gamma}^2({\bf k})
\frac{1}{i\varepsilon_n \!-\! \xi({\bf k} \!+\! \frac{1}{2}{\bf q})}
\nonumber  \\ && \times
\frac{1}{- i\varepsilon_n - i\omega_l-\xi(-{\bf k} + \frac{1}{2}{\bf q})} ,
\quad 
\end{eqnarray} 
Only $d\gamma$-wave SC fluctuation are considered in this paper because $T_c$ of
$d\gamma$ wave are definitely much higher than $T_c$ of other
waves.\cite{OhSC1,OhSC2}
An effective CBO susceptibility is similarly given by
\begin{equation}\label{EqCBOsus}
\chi_{\Gamma}^{\mbox{\tiny (CBO)}}(i\omega_l,{\bf q}) =
\frac{\pi_{\Gamma}^{\mbox{\tiny (CBO)}}(i\omega_l,{\bf q})}
{1 + \frac{3}{4} I_1^* \tilde{W}_s^2
\pi_{\Gamma}^{\mbox{\tiny (CBO)}}(i\omega_l,{\bf q})},
\end{equation}
for $\Gamma=s$, $p$ and $d\gamma$ waves, with 
\begin{eqnarray}
\pi^{\mbox{\tiny (CBO)}}_{\Gamma}(i\omega_l,{\bf q}) &=&
- \frac{k_B T}{N}\sum_{n{\bf k}} 
\eta_{\Gamma}^2({\bf k})
\frac{1}{i\varepsilon_n \!-\! \xi({\bf k} \!-\! \frac{1}{2}{\bf q})}
\nonumber \\ && \times
\frac{1}{i\varepsilon_n+i\omega_l-\xi({\bf k} + \frac{1}{2}{\bf q})} .
\end{eqnarray}
The form factors of $p$ waves are defined by
\begin{equation}
\eta_{x}({\bf k})  = \sqrt{2}\sin(k_xa), \quad
\eta_{y}({\bf k})  = \sqrt{2}\sin(k_ya).
\end{equation}

According to Eq.~(\ref{EqSCsus}),
$T_c$ of $d\gamma$ wave superconductivity are given by
\begin{equation}\label{EqTc}
1 + \mbox{$\frac{3}{4}$} I_1^* \tilde{W}_s^2
\pi_{d}^{\mbox{\tiny (SC)}}(0,0)=0 .
\end{equation}
It was shown in the previous papers\cite{OhSC1,OhSC2} that
\begin{equation}\label{EqI-Star}
\mbox{$\frac{3}{4}$}|I_1^*|\tilde{W}_s^2  \simeq 100\mbox{~meV}
\end{equation}
is needed  in order to explain observed $T_c$.
The Wilson ratio is as large as $\tilde{W}_s\simeq 2$ in SSA,
and the superexchange interaction is as strong as
$J=-(100\mbox{--}150)~\mbox{meV}$
in actual cuprate oxide superconductors. Then, it follows that
$\mbox{$\frac{3}{4}$}|I_1^*|\tilde{W}_s^2  \agt 400\mbox{~meV}$;
$|I_1^*|>|J|$. The theory published in 1987 has a drawback that it gives too
high theoretical
$T_c$ to explain observed $T_c$.\cite{OhSC1,OhSC2} The mass enhancement
factor and the effective three-point vertex function are
renormalized by AF, SC and CBO fluctuations, so that $\tilde{W}_s$ that is
the effective three-point vertex function divided by the mass enhancement
factor is also renormalized by the fluctuations.
We argued in the previous paper\cite{OhSupJ2} that
the renormalization of $\tilde{W}_s$ is substantial so that
$\tilde{W}_s \simeq 1$ or $\tilde{W}_s \alt 1$;
a phenomenological argument also implies that
$\tilde{W}_s = 0.7 \mbox{--} 1$
had better been used in order to explain quantitatively $T_c$ and $T$-linear
resistivities. For example, Eq.~(\ref{EqI-Star}) can only be
satisfied for such small 
$\tilde{W}_s$.  We follow this argument.
Taking the renormalization into account, we regard 
$\tilde{W}_s$ as a phenomenological parameter; we  assume 
\begin{equation}
\tilde{W}_s = 1 
\end{equation}
in this paper.
\vspace{1cm}

\subsection{Renormalization of phonons}

\begin{figure*}
\centerline{\hspace*{0.cm}
\includegraphics[width=18.5cm]{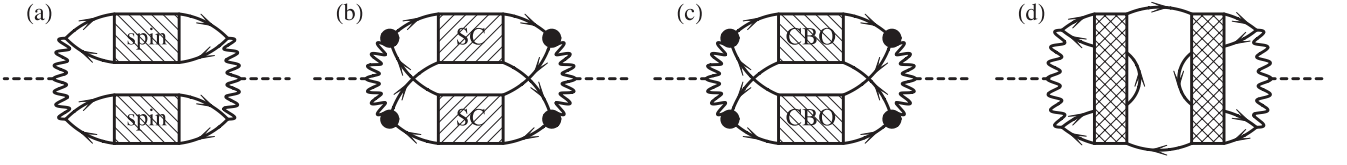}}
\caption[2]{
Four processes of the renormalization of phonons.   A solid line
stands for an electron, a broken line for a phonon, a wavy line for the
superexchange interaction $J$, and a solid circle for the effective vertex
function $\tilde{\phi}_s$.  Hatched parts in  Figs.~\ref{fig}(a), (b) and (c)
stand for AF (spin), SC and CBO fluctuations, respectively, and a hatched
part, including an internal electron line, in  Figs.~\ref{fig}(d) for the
vertex function $Z(i\varepsilon_n,i\omega_l; {\bf k},{\bf q})$. The
contribution from  Figs.~\ref{fig}(d) is larger than those of
Figs.~\ref{fig}(a), (b) and (c); two {\it fluctuation-lines} with different 
{\bf q} or two susceptibilities of different {\bf q} appear in the
convolution form in Figs.~\ref{fig}(a), (b) and (c) so that their
contributions can  be large only when fluctuations are developed in a wide
region of the momentum space.
 }
\label{fig}
\end{figure*}

The Green function for
phonons  is given by
\begin{equation}
D_\lambda(i\omega_l,{\bf q})=
\frac{2 \omega_{\lambda{\bf q}} } 
{ (i\omega_l)^2 \!-\! \omega_{{\bf q}\lambda}^2
\!+\! 2 \omega_{\lambda{\bf q}}\Delta\omega_\lambda(i\omega_l,{\bf q})
},
\end{equation}
with
\begin{equation}\label{EqSoftOmega}
\Delta\omega_\lambda(i\omega_l,{\bf q}) = - 
\frac{\hbar^2}{2 M_p \omega_{\lambda{\bf q}} } 
S (i\omega_l,{\bf q}) .
\end{equation}
Because phonons are renormalized by AF, SC and CBO fluctuations as well as
charge density fluctuations, 
we consider four processes shown in Fig.~\ref{fig}: 
$S(i\omega_l,{\bf q}) = S_{s}(i\omega_l,{\bf q})  
+ S_{\mbox{\tiny SC}}(i\omega_l,{\bf q}) +
S_{\mbox{\tiny CBO}}(i\omega_l,{\bf q})  
+ S_{c}(i\omega_l,{\bf q}) $. 
The expression~(\ref{EqTwoSpin}) for 
${\cal P}_\Gamma({\bf q})$ is convenient
in treating couplings with AF fluctuations. Other expressions, which
are convenient in treating couplings with SC and CBO
fluctuations, are shown in Appendix.
When  Eqs.~(\ref{EqEl-Ph-S}), (\ref{EqEl-Ph-CBO2}), and (\ref{EqEl-Ph-SC2})
 are made use of and only the parts of
$\Gamma=s$ in Eqs.~(\ref{EqElPhP}) and (\ref{EqElPhD}) are considered,
 it follows that 
\begin{widetext}
\begin{equation}\label{EqS-SDW}
S_{s}(i\omega_l,{\bf q}) = \frac{3}{4^2}  
Y_{\lambda}^2({\bf q}) 
\frac{k_B T}{N} \sum_{l^\prime {\bf q}^\prime} 
\eta_{s}^2 ({\bf q}^\prime)
\chi_s\!\left(i\omega_l+i\omega_{l^\prime}, 
{\bf q}^\prime + \mbox{$\frac{1}{2}$}{\bf q}\right)
\chi_s\!\left(-i\omega_{l^\prime}, 
-{\bf q}^\prime + \mbox{$\frac{1}{2}$}{\bf q}\right),
\end{equation}
\begin{eqnarray}\label{EqS-SC}
S_{\mbox{\tiny SC}}(i\omega_l,{\bf q}) &=&
\frac{3^2}{4^3} \tilde{W}_s^4 Y_{\lambda}^2({\bf q}) 
\frac{k_B T}{N}  \sum_{l^\prime{\bf q}^\prime} 
\biggl[
\chi_{d}^{\mbox{\tiny SC}}\!\left(i\omega_l+i\omega_{l^\prime}, 
{\bf q}^\prime + \mbox{$\frac{1}{2}$}{\bf q}\right)
\chi_{d}^{\mbox{\tiny SC}}\!\left(-i\omega_{l^\prime}, 
-{\bf q}^\prime + \mbox{$\frac{1}{2}$}{\bf q}\right) 
\nonumber \\ && \quad 
- \pi_{d}^{\mbox{\tiny SC}}\!\left(i\omega_l+i\omega_{l^\prime}, 
{\bf q}^\prime + \mbox{$\frac{1}{2}$}{\bf q}\right)
\pi_{d}^{\mbox{\tiny SC}}\!\left(-i\omega_{l^\prime}, 
-{\bf q}^\prime + \mbox{$\frac{1}{2}$}{\bf q}\right)
\biggr],
\end{eqnarray}
\begin{eqnarray}\label{EqS-CBO}
S_{\mbox{\tiny CBO}}(i\omega_l,{\bf q}) &=&
\frac{3^2}{4^3} \tilde{W}_s^4 Y_{\lambda}^2({\bf q}) 
\sum_{\Gamma}
\frac{k_B T}{N} \sum_{l^\prime{\bf q}^\prime}
\biggl[
\chi_{\Gamma}^{\mbox{\tiny CBO}}\!\left(i\omega_l+i\omega_{l^\prime}, 
{\bf q}^\prime + \mbox{$\frac{1}{2}$}{\bf q}\right)
\chi_{\Gamma}^{\mbox{\tiny CBO}}\!\left(-i\omega_{l^\prime}, 
-{\bf q}^\prime + \mbox{$\frac{1}{2}$}{\bf q}\right) 
\nonumber \\ && \quad 
- \pi_{d}^{\mbox{\tiny CBO}}\!\left(i\omega_l+i\omega_{l^\prime}, 
{\bf q}^\prime + \mbox{$\frac{1}{2}$}{\bf q}\right)
\pi_{d}^{\mbox{\tiny CBO}}\!\left(-i\omega_{l^\prime}, 
-{\bf q}^\prime + \mbox{$\frac{1}{2}$}{\bf q}\right)
\biggr],
\end{eqnarray}
\begin{equation}\label{EqS-C}
S_{c}(i\omega_l,{\bf q}) =
- \frac{3^2}{4^2} \tilde{W}_s^4  Y_{\lambda}^2({\bf q})
\frac{k_B T}{N}  \sum_{n{\bf k}\sigma} 
Z^2(i\varepsilon_n,i \omega_l; {\bf k},{\bf q})
\frac{1}{i\varepsilon_n \!-\! \xi({\bf k}) }
\frac{1}{i\varepsilon_n \!+\! i \omega_l \!-\!
\xi({\bf k}\!+\! {\bf q} ) } ,
\end{equation}
with 
\begin{equation} 
Y_{\lambda}({\bf q}) = 
\bar{\eta}_s({\bf q}) \left[C_p v_{p,\lambda{\bf q}} 
\eta_{s} \left(\mbox{$\frac{1}{2}{\bf q}$}\right)   
+  C_d v_{d,\lambda{\bf q}}\sqrt{M_p/M_d} \right].
\end{equation}
In Eq.~(\ref{EqS-C}),
$Z(i\varepsilon_n,i
\omega_l; {\bf k},{\bf q})$ is the vertex function in the
charge channel. It is also enhanced by AF, SC and CBO fluctuations; 
$Z(i\varepsilon_n,i \omega_l; {\bf k},{\bf q}) = 
Z_{s}(i\varepsilon_n,i \omega_l; {\bf k},{\bf q}) + 
Z_{\mbox{\tiny SC}}(i\varepsilon_n,i \omega_l; {\bf k},{\bf q}) +
Z_{\mbox{\tiny CBO}}(i\varepsilon_n,i \omega_l; {\bf k},{\bf q})  +\cdots$, 
with 
\begin{equation}\label{EqZS}
Z_{s}(i\varepsilon_n,i \omega_l; {\bf k},{\bf q}) =
\frac{k_BT}{N} \sum_{l^\prime{\bf q}^\prime} 
\eta_{s}({\bf q}^\prime) 
\frac{
K_s\left(i\omega_{l^\prime}, 
{\bf q}^\prime + \mbox{$\frac{1}{2}$}{\bf q} \right) 
K_s\left(-i\omega_{l^\prime} + i \omega_l, 
-{\bf q}^\prime + \mbox{$\frac{1}{2}$}{\bf q} \right)
}{ 
i\varepsilon_n + i \omega_l^\prime 
- \xi({\bf k} +  {\bf q}^\prime + \mbox{$\frac{1}{2}$}{\bf q} ) 
} ,
\end{equation}
\begin{eqnarray}\label{EqZSC}
Z_{\mbox{\tiny SC}}(i\varepsilon_n,i \omega_l; {\bf k},{\bf q}) &=&
\frac{1}{2} 
\frac{k_BT}{N} \sum_{l^\prime{\bf q}^\prime} 
\eta_{d}\left({\bf k}- \mbox{$\frac1{2}$}{\bf q}^\prime 
+ \mbox{$\frac1{4}$}{\bf q}\right)
\eta_{d}\left({\bf k}- \mbox{$\frac1{2}$}{\bf q}^\prime 
+ \mbox{$\frac{3}{4}$}{\bf q}\right)
\nonumber \\ && \quad \times 
\frac{
K_{d}^{\mbox{\tiny SC}}
\left(i\omega_{l^\prime}, 
{\bf q}^\prime - \mbox{$\frac{1}{2}$}{\bf q} \right)
K_{d}^{\mbox{\tiny SC}}
\left(i\omega_{l^\prime} + i \omega_l, 
{\bf q}^\prime + \mbox{$\frac{1}{2}$}{\bf q} \right)
}{ 
- i\varepsilon_n + i \omega_l^\prime 
- \xi(-{\bf k} + {\bf q}^\prime - \mbox{$\frac{1}{2}$}{\bf q} ) },
\end{eqnarray}
\begin{eqnarray}\label{EqZCBO}
Z_{\mbox{\tiny CBO}}(i\varepsilon_n,i \omega_l; {\bf k},{\bf q})  &=&
- \frac{1}{2}
\frac{k_BT}{N} \!\sum_{l^\prime{\bf q}^\prime} 
\sum_{\Gamma} 
\eta_{\Gamma}\left({\bf k}+ \mbox{$\frac1{2}$}{\bf q}^\prime 
+ \mbox{$\frac1{2}$}{\bf q}\right)
\eta_{\Gamma}\left({\bf k}+ \mbox{$\frac1{2}$}{\bf q}^\prime 
+ \mbox{$\frac{3}{4}$}{\bf q}\right)
\nonumber \\ && \quad \times \frac{
K_{\Gamma}^{\mbox{\tiny CBO}}
\left(i\omega_{l^\prime} , 
{\bf q}^\prime \!+\! \mbox{$\frac{1}{2}$}{\bf q} \right) 
K_{\Gamma}^{\mbox{\tiny CBO}}
\left(-i\omega_{l^\prime} + i \omega_l, 
-{\bf q}^\prime + \mbox{$\frac{1}{2}$}{\bf q} \right) 
}{
i\varepsilon_n + i \omega_l^\prime 
- \xi({\bf k} +  {\bf q}^\prime + \mbox{$\frac{1}{2}$}{\bf q} ) },
\end{eqnarray}
with
\end{widetext}
\begin{equation}\label{EqKS}
 K_s(i\omega_l,{\bf q}) =
\frac{1}{1 \!-\! \frac{1}{4} I(i\omega_l,{\bf q})
\tilde{\chi}_s(i\omega_l)}, 
\end{equation}
\vspace{-0.3cm}
\begin{equation}\label{EqKSC}
K_{d}^{\mbox{\tiny SC}} (i\omega_l,{\bf q}) =
\frac1{1 \!+\! \frac{3}{4} I_1^* \tilde{W}_s^2
\pi_{d}^{\mbox{\tiny SC}}(i\omega_l,{\bf q}) } -1,
\end{equation}
\vspace{-0.3cm}
\begin{equation}\label{EqKCBO}
K_{\Gamma}^{\mbox{\tiny CBO}} (i\omega_l,{\bf q}) =
\frac1{1 \!+\! \frac{3}{4} I_1^* \tilde{W}_s^2
\pi_{\Gamma}^{\mbox{\tiny CBO}}(i\omega_l,{\bf q}) } -1. 
\end{equation}
Here, Eqs.~(\ref{EqEl-Ph-S}), (\ref{EqEl-Ph-CBO2}), and (\ref{EqEl-Ph-SC2})
are also made use of;
zero-th order terms in $I_1^*$ are
subtracted in Eqs.~(\ref{EqS-SC}), (\ref{EqS-CBO}), (\ref{EqKSC}) and
(\ref{EqKCBO}) to avoid any double counting.

\section{Application to Cuprate Oxide Superconductors}
\subsection{Softening of phonons}

Because it is not a purpose of this paper to study phonon modes themselves,
we study only issues that can be clarified without calculating them. 
First, we consider the softening of the so called half breathing mode of O
ions with ${\bf Q}_X =(\pm\pi/a,0)$ or $(0,\pm\pi/a)$.  Because 
\begin{equation}
v_{d,\lambda{\bf Q}_X}=0, \quad 
v_{p,\lambda{\bf Q}_X} =1 ,
\end{equation}
we have to consider only the electron-phonon interaction given by
Eq.~(\ref{EqElPhP}).

Because doped holes mainly go into O ions, one may argue
that $3d$ levels of Cu ions are deeper $2p$ levels of O ions, 
$\epsilon_d <\epsilon_p$. However, this argument disagrees with what band
calculations\cite{band1,band2,band3} predict,
$\epsilon_{d}-\epsilon_{p} \simeq 1\mbox{~eV}$. The
observation does not necessarily mean that $\epsilon_d <\epsilon_p$, but it
simply means that the local charge susceptibility of $3d$ electrons is much
smaller than  that of $2p$ electrons, as is discussed in Introduction. Because
it is unlikely that band calculations give such a bad prediction on relative
positions between the $3d$ and $2p$ levels, we follow what band
calculations predict. When we use  
\begin{equation}
V = 1.6\mbox{~eV}, \quad 
\epsilon_{d}-\epsilon_{p} = 1\mbox{~eV}, \quad  
U = 5\mbox{~eV}, 
\end{equation}
following the previous paper,\cite{OhSupJ1} Eq.~(\ref{EqSuperJ}) gives
$J = - 0.27~\mbox{eV}$. 
This is about twice as large as the experimental one of
$J = - (0.10\mbox{--}0.15)~\mbox{eV}$.
This discrepancy is resolved when nonzero bandwidths of the lower and upper
Hubbard band are considered.\cite{OhSupJ1}
Because of this reduction, we assume a half of $C_p$ given by
Eq.(\ref{EqCp}):
$C_p = 0.1 \times A_d$ instead of $C_p = 0.22 \times A_d $. 
When we take $A_d\simeq 5$~eV/\AA, it follows that
\begin{equation}
C_p \simeq 0.5 \mbox{~eV}/\mbox{\AA}, \quad 
Y_\lambda({\bf Q}_X) \simeq 1 \mbox{~eV}/\mbox{\AA} .
\end{equation}

Two susceptibilities appear in the  convolution form in
Eqs.~(\ref{EqS-SDW}), (\ref{EqS-SC}) and (\ref{EqS-CBO}).
Unless AF, SC and CBO fluctuations are developed in a wide region of the
momentum space, the convolutions cannot be large. We assume that
$S_{c}(i\omega_l,{\bf q})$ given by Eq.~(\ref{EqS-C}) is dominant
in  $S(i\omega_l,{\bf q})$. When  only 
$S_c(i\omega_l,{\bf q})$ is considered, 
\begin{equation}\label{EqSApprox}
S(i\omega_l,{\bf q}) \simeq
\frac{3^2}{4^2} \tilde{W}_s^4 
\left<Z^2 \right>
\frac{Y_\lambda^2({\bf Q}_X)} {2 k_BT_K} , 
\end{equation}
with $\left<Z^2 \right>$
an average of
$Z^2(i\varepsilon_n, i\omega_l; {\bf k}, {\bf q})$.
When we assume $\tilde{W}_s=1$,
the softening at X point is given by
\begin{equation}
\!\Delta\omega_\lambda(\omega_{\lambda{\bf q}_X},{\bf Q}_X)
\simeq - 0.01 
\left<Z^2\right>
\frac{(10^3~\mbox{meV})^2}{\omega_{\lambda{\bf Q}_X} k_BT_K}
~\mbox{meV}.  
\end{equation}
 When we put
\begin{equation} 
\omega_{\lambda{\bf Q}_X} \simeq k_B T_K \simeq  10^2 \mbox{~meV},
\end{equation}
it follows that 
\begin{equation}
\Delta\omega_\lambda(\omega_{\lambda{\bf q}_X},{\bf Q}_X)
\simeq \left<Z^2\right>~\mbox{meV}.
\end{equation}
When AF, SC and CBO fluctuations are not developed,
$\left<Z^2\right> \alt 1$ and 
the softening must be very small.
When they are well developed so that
$\left<Z^2\right>$ might be as large as 
$\left<Z^2\right> \simeq 10$,
we can  explain the observed softening as large as
\begin{equation}\label{EqObsSoft}
\Delta\omega_\lambda(\omega_{\lambda{\bf q}_X},{\bf Q}_X)
\simeq 10 ~\mbox{meV}.
\end{equation}
When other contributions are considered in addition to
$S_{c}(i\omega_l,{\bf q})$, $\left<Z^2\right>$
can be smaller than 10 to explain the observed softening.

No softening occurs for ${\bf q}=0$
because $\bar{\eta}_s (0) =0$.
When ${\bf q}$ goes from $\Gamma$ point to $X$ point, the softening
must increase first but it is unlikely that the  softening is the largest at
$X$ point.
Because $v_{d,\lambda{\bf Q}_X} =0$,  the electron-phonon interaction
described by  Eq.~(\ref{EqElPhD}) vanishes. 
This implies that the softening cannot be the largest at $X$ point 
along $\Gamma$-$X$ line. 
In actual, several experimental data imply
that the softening is the largest for ${\bf q}$ a little different from 
${\bf Q}_X$ along $\Gamma$-$X$ line. \cite{Pint1,Braden} 
 
No softening cannot occur either at $M$
point or for the breathing mode of O ions with
${\bf Q}_M =(\pm\pi/a,\pm\pi/a)$ because
$v_{d,\lambda{\bf Q}_M}=0$ and $\eta_s(\mbox{$\frac1{2}$}{\bf Q}_M)=0$.
It is interesting to confirm this prediction.

\subsection{Kinks in the quasiparticle dispersion}

A process corresponding to Fig.~1(d) renormalizes quasiparticles.
The self-energy correction is given by 
\begin{eqnarray}
\frac{1}{\tilde{\phi}_\gamma }\Delta \Sigma (i\varepsilon_n, {\bf k}) 
\!\! &=& \!\!
- \frac{k_B T}{N} \sum_{\lambda l{\bf q}} 
g_\lambda ^2(i\varepsilon_n,i \omega_l;{\bf k},{\bf q})
D_\lambda(i\omega_l,{\bf q})
\nonumber \\ && \times
\frac{1}{i\varepsilon_n+i\omega_l
-\xi({\bf k}+{\bf q})} ,
\end{eqnarray}
with
\begin{eqnarray}\label{EqG}
g_\lambda(i\varepsilon_n,i \omega_l; {\bf k},{\bf q}) &=&
 \frac{\hbar}{\sqrt{2M_p \omega_{\lambda{\bf q}}} }
\frac{3}{4}\tilde{W}_s^2 Y_\lambda({\bf q})
\nonumber \\ && \quad \times 
Z(i\varepsilon_n,i \omega_l; {\bf k},{\bf q}) . \qquad 
\end{eqnarray}
It is likely  that the contribution of Fig.~\ref{fig}(d) dominate those
of the other three, Figs.~\ref{fig}(a)--(c).  In such a case, 
\begin{equation}
g_\lambda(i\varepsilon_n,i \omega_l; {\bf k},{\bf Q}_X) \simeq
\sqrt{ 2 k_B T_K |\Delta\omega_\lambda
(\omega_{\lambda{\bf q}},{\bf Q}_X)|} .
\end{equation}
Here, Eqs.(\ref{EqSoftOmega}) and (\ref{EqSApprox}) are made use of.
When the experimental value (\ref{EqObsSoft}) is used, we obtain 
\begin{equation}\label{EqG-X}
g_\lambda(i\varepsilon_n,i \omega_l; {\bf k},{\bf Q}_X) \simeq
45 \mbox{~meV}.
\end{equation}
This is large enough for optical phonons to cause kinks in the
quasiparticle dispersion. 

Two types of kinks are observed.\cite{sato} The renormalization by phonons
can explain one type of kinks observed in both normal and SC phases. However,
it is difficult to explain the other type of kinks observed only in SC
phases; low-energy AF and SC fluctuations are suppressed when SC gaps open.

\subsection{Cooper-pair interaction}

The phonon-mediated pairing interaction is given by
\begin{equation}
V_{ph}({\bf q};{\bf k}) =
- 2g_\lambda ^2(0, 0;{\bf k},{\bf q})/\omega_{\lambda{\bf q}}.
\end{equation}
Its average over ${\bf k}$ on the Fermi
surface is expanded in such a way that
\begin{equation}
\left<V_{ph}({\bf q};{\bf k})\right> = 
V_{0} + 2 V_{1}\eta_s({\bf q})
+  2 V_{2}\eta_{s2}({\bf q}) + \cdots.
\end{equation}
No softening at $\Gamma$ and $M$ points implies
$\left<V_{ph}(0;{\bf k})\right>= 0$ and
$\left<V_{ph}({\bf Q}_M;{\bf k})\right>= 0$, so that
\begin{equation}
V_0+2V_1+2V_2\simeq 0, \quad
V_0-2V_1+2V_2\simeq 0.
\end{equation}
Because of Eq.~(\ref{EqG-X}), 
$\left<V_{ph}({\bf Q}_X;{\bf k})\right>\simeq - 40~\mbox{meV}$ or
\begin{equation}
V_0-2V_2\simeq - 40~\mbox{meV}.
\end{equation}
Then, we obtain 
\begin{equation}\label{EqPhPairInt}
V_0\simeq - 20\mbox{~meV}, \quad V_1\simeq 0\mbox{~meV},  
\quad V_2\simeq 10\mbox{~meV}. 
\end{equation}

The interaction
$V_1$ between nearest neighbors should be included in addition to
$\frac{3}{4}I_1^*\tilde{W}_s^2$ in the theory of $d\gamma$-wave high-$T_c$
superconductivity.  When Eq.~(\ref{EqTc}) is extended to include $V_1$, 
$T_c$ are determined by
\begin{equation}
1 + \left(\mbox{$\frac{3}{4}$} I_1^* \tilde{W}_s^2 + V_{1}\right)
\pi_{d}^{\mbox{\tiny (SC)}}(0,0)=0 .
\end{equation}
Although the {\bf q} dependence of 
$\left<V_{ph}({\bf q};{\bf k})\right>$ in the whole Brillouin zone is
necessary to estimate $V_1$ accurately, we can conclude that
$|V_1|$ must be much smaller than 
$\frac{3}{4}|I_1^*|\tilde{W}_s^2  \simeq 100\mbox{~meV}$,
which is needed in order to explain observed $T_c$.\cite{OhSC1,OhSC2}.

There are various branches of phonon modes beside the mode discussed
above. The virtual exchange of phonons
that do not become soft
cannot give a significant pairing interaction.

\section{Discussion}

Following Barnes,\cite{barnes} we can map the $t$-$J$ model to the so called
auxiliary-particle $t$-$J$ model:
$\bar{\cal H}_{t\mbox{-}J}={\cal P}^{-1}\bar{\cal H}\hspace{1pt}{\cal P}$,
with
\begin{eqnarray}\label{auxiliary-model}
\bar{\cal H} &=&   
\delta \sum_{i} \left(\! e_{i}^{\dag} e_{i} 
+ c_{i\uparrow}^{\dag} c_{i\uparrow} 
+ c_{i\downarrow}^{\dag} c_{i\downarrow} -1 \right) 
+ \epsilon_d  \sum_{i\sigma}  c_{i\sigma}^\dag c_{i\sigma} 
\nonumber \\ && \quad 
- \sum_{ij\sigma} t_{ij} c_{i\sigma}^\dag
e_{i} c_{j\sigma}e_{j}^\dag 
-  \frac1{2}J 
\sum_{\left<ij\right>} (\bar{\bf S}_i \cdot \bar{\bf S}_j) , \quad 
\end{eqnarray}
with 
$\delta$ being an arbitrary constant,\cite{not-unique} and 
\begin{equation} 
\bar{\bf S}_i = \frac1{2} \sum_{\alpha\beta}  \left(
\sigma_x^{\alpha\beta} \!, \sigma_y^{\alpha\beta} \!,
\sigma_z^{\alpha\beta} \right) c_{i\alpha}^\dagger c_{i\beta}.
\end{equation}
Two kinds of auxiliary particles, which correspond to empty and
occupied sites in the original $t$-$J$ model, are introduced; 
$e_i^\dag$ and $c_{i\sigma}^\dag$ are their creation operators.
We call them $e$ and $c$ particles in this paper.
The projection operator ${\cal P}$ restricts the Hilbert space  within 
\begin{equation}\label{EqQ-rest}
Q_{i} \equiv  e_{i}^{\dag} e_{i} + c_{i\uparrow}^{\dag} c_{i\uparrow}
+ c_{i\downarrow}^{\dag} c_{i\downarrow} =1
\end{equation}
for any $i$ site; no empty or multiply occupied
sites are allowed. This restriction is guaranteed by the conservation of the
number of auxiliary particles such as 
\begin{equation}
\left[\bar{\cal H}, Q_{i}\right]=0
\end{equation} 
for any $i$, or local gauge symmetry. This symmetry is inherent in the
auxiliary-particle model. Local gauge symmetry can never be
broken.\cite{elitzur} Therefore, no single auxiliary particle can be added or
removed, or no single-particle excitation of auxiliary particles is allowed.
Auxiliary particles themselves are never itinerant but are localized; pair
excitations of auxiliary particles are itinerant.   Fermionic pair
excitations of auxiliary particles correspond to electrons in $t$-$J$ model.
Two ways of statistics are possible: fermionic $e$ and bosonic $c$ particles,
and fermionic $c$ and bosonic $e$ particles. The model with bosonic $e$ and
fermionic $c$ particles  is often called the slave-boson $t$-$J$
model.\cite{ComSlave} The mean-field (MF) theory for  $\bar{\cal H}$, instead
of ${\cal P}^{-1}\bar{\cal H}\hspace{1pt}{\cal P}$, and its more or less
improved theories, which include gauge fluctuations,  assume the breaking of
local gauge symmetry, and they treat single-particle excitations of itinerant
auxiliary particles. Such theories are never relevant to study dynamics of
electrons in the original $t$-$J$ model; condensation energies derived in
these treatments are consistent with Gutzwiller's theory and are reliable. A
more precise discussion on this issue can be found in Appendix of
Ref.~\onlinecite{OhSupJ1}.  Although an apparently  similar theoretical
development to that of this paper is possible when one starts from the MF
approximation for the slave-boson $t$-$J$ model, it is  physically and
essentially  different from the theory of Kondo lattices; states considered
in the two theories are of totally different symmetry from each other.
Therefore, we should abstain from comparing results based on the theory of
Kondo lattices with those of the MF theory of the slave-boson $t$-$J$
model.\cite{slaveP}

Note that $\tilde{\phi}_c$ and 
$1/\tilde{\phi}_\gamma$ are small parameters in the vicinity of the Mott-Hubbard
transition.  What are considered in this
paper are of leading order in both  $\tilde{\phi}_c$ and 
$1/\tilde{\phi}_\gamma$, that is,  order of
$(\tilde{\phi}_c)^0(1/\tilde{\phi}_\gamma)^0$.

There are two other types of electron-phonon interactions:
the modulation of $3d$-electron
levels, $\epsilon_d$,  and that of the transfer integrals, $t_{ij}$. 
The conventional one arising from the modulation of
$\epsilon_d$, which can directly couples with charge fluctuations, gives
renormalization effects higher order in 
$\tilde{\phi}_c$ and $1/\tilde{\phi}_\gamma$, so that its effects must be
very small. The electron-phonon interaction
arising from the modulation of $t_{ij}$  gives renormalization effects higher
order in  $1/\tilde{\phi}_\gamma$, so that 
their effects of the electron-phonon interaction must be 
$1/\tilde{\phi}_\gamma^2$ times as small as those studied in this paper;
we expect that coupling constants for $t_{ij}$, which
correspond to $A_d$, $A_p$ and $A_V$ of this paper, are of the same oder of
magnitude as $A_V$. 
Then, we ignore both of them in this paper.

There are also pieces of experimental evidence that the electron-phonon
interaction arising from the modulation of
$\epsilon_d$ or $t_{ij}$ by phonons is irrelevant in cuprates. Hwang,
Timusk and Gu investigated life-time widths of quasiparticles instead of
kinks; life-time widths depend on temperature and dopings.\cite{timusk} 
Their observation implies that kinks are large only in metallic cuprates
where AF and SC fluctuations are well developed. The softening of phonons is
also large in such metallic cuprates; no significant softening is observed in
over-doped cuprates.\cite{fukuda} It is difficult to explain these
observations in terms of the electron-phonon interaction arising from the
modulation of
$\epsilon_d$ or $t_{ij}$ by phonons.   On the other hand,  these observations
are pieces of evidence that the electron-phonon interaction that can couple
directly with AF and SC channels is relevant and its enhancement by AF and SC
fluctuations in metallic phase is crucial.

In insulating phases, only the contributions from
Eq.~(\ref{EqS-SDW}) remains but those from
Eqs.~(\ref{EqS-SC})-(\ref{EqS-C}) vanish. Then, we cannot expect
significant softening of phonons. This is also consistent with
experiment.\cite{fukuda}

Various physical properties are different or asymmetric between hole-doped
and electron-doped cuprates. Within the theoretical framework of this paper,
hole-doped and electron-doped cuprates must be, in essence, similar to each
other. Phenomenologically, AF and SC fluctuations are relatively more
developed in hole-doped cuprates than they are in electron-doped cuprates. If
the asymmetry of the fluctuations can be explained, we can explain that of
phonon properties. It is pointed out in another paper\cite{OhDisorder} that
the asymmetry of disorder can play a crucial role in the asymmetry between
hole-doped and electron-doped cuprates.

In cuprate oxide superconductors, the exchange interaction arising from the
virtual exchange of pair excitations of quasiparticles is less effective than
the superexchange interaction; the pairing interaction arising from
phonons can play no significant role. Then, the main pairing interaction
$I_1^*$ must arise from the superexchange interaction, which is enhanced by
spin fluctuations.

As is discussed in Introduction, the superexchange interaction arises from the
virtual exchange of pair excitations of electrons across the lower and upper
Hubbard bands.  As is shown in Eqs.~(\ref{EqIs*1}) and (\ref{EqIs*2}), the
spin-fluctuation mediated pairing interaction is essentially the same as the
superexchange interaction if 
high-energy spin fluctuations, whose energies are as large as the Hubbard
onsite repulsion $U$, are properly included.  However, it is  physically
different from the superexchange interaction if only low-energy spin
fluctuations are included.

The SSA is rigorous for Landau's Fermi-liquid states in infinite
dimensions,\cite{metzner} so that the theory of Kondo lattices 
can be regarded as a $1/d$ expansion theory, with $d$ being the spatial
dimensionality.  One may suspect that the $1/d$ expansion theory cannot be
applied to quasi-two dimensional cuprates. Any perturbative theory relies on
the analytical continuity;\cite{anderson} a perturbed state must be of the
symmetry as an unperturbed state is. Normal states in over-doped or
optimal-doped cuprates are certainly Landau's normal Fermi liquids. Because
there is no evidence that any symmetric change occurs between normal states
in over-doped or optimal-doped cuprates and exotic  normal states in
under-doped cuprates, we can argue that the analytical continuity holds so
that the $1/d$ expansion theory or the theory of Kondo lattices can be
applied to exotic normal states in cuprates.

\section{Conclusion}

The electron-phonon interaction arising from the modulation of the
superexchange interaction by  phonons is  relevant for strongly correlated
electron liquids in the vicinity of the Mott-Hubbard transition.  
It is shown with the help of the theory of Kondo lattices 
that it can be enhanced by spin, superconducting, and charge bond-order
fluctuations as well as charge fluctuations.   The enhanced
electron-phonon interaction is  responsible for not only the softening of
phonons but also kinks in the dispersion relation of quasiparticles in
cuprate oxide high-temperature  superconductors. However, it can never be the
main Cooper-pair interaction.  The main one must be the superexchange
interaction.

\begin{acknowledgments}
This work is supported by a Grant-in-Aid for Scientific Research (C) Grant
No.~13640342 from the Japan Society for the Promotion of Science. 
\end{acknowledgments}

\appendix
\section{Various expressions for the electron-phonon interaction}

Equation~(\ref{EqTwoSpin}) is also written in another form:
\begin{widetext}
\begin{equation}\label{EqEl-Ph-S}
{\cal P}_\Gamma({\bf q})  = 
\frac1{2N} \sum_{{\bf k}{\bf p}{\bf q}^\prime} 
\sum_{\alpha\beta\gamma\delta}
\eta_{\Gamma}({\bf q}^\prime) 
\left({\bf s}^{\alpha\beta} \!\cdot {\bf s}^{\gamma\delta}\right)
a_{({\bf k} + \frac1{2}{\bf q}^\prime + \frac1{2}{\bf q}) \alpha}^\dag 
a_{({\bf k} - \frac1{2}{\bf q}^\prime) \beta}
a_{({\bf p} - \frac1{2}{\bf q}^\prime + \frac1{2}{\bf q})\gamma}^\dag
a_{({\bf p} + \frac1{2}{\bf q}^\prime)\delta} .
\end{equation} 
This expression is useful to obtain Eqs.~(\ref{EqS-SDW}) and (\ref{EqZS}).
When we replace variables in Eq.~(\ref{EqEl-Ph-S}) in such a way that
${\bf k} + \frac1{2} {\bf q}^\prime = {\bf k}_1 + \frac1{2}{\bf q}_1$, 
${\bf k} - \frac1{2} {\bf q}^\prime = {\bf p}_1 + \frac1{2} {\bf q}_1$,
and
${\bf p} + \frac1{2} {\bf q}^\prime = {\bf k}_1 - \frac1{2}{\bf q}_1$, 
Eq.~(\ref{EqEl-Ph-S}) turns out to
\begin{equation}\label{EqEl-Ph-CBO}
{\cal P}_\Gamma ({\bf q}) =
\frac1{2N}\sum_{{\bf k}_1{\bf p}_1{\bf q}_1} 
\sum_{\alpha\beta\gamma\delta}
\eta_{\Gamma}({\bf k}_1-{\bf p}_1) 
\left({\bf s}^{\alpha\beta} \!\cdot {\bf s}^{\gamma\delta}\right)
a_{({\bf k}_1 +  \frac1{2}{\bf q}_1 +  \frac1{2}{\bf q})\alpha}^\dag 
a_{({\bf p}_1 + \frac1{2}{\bf q}_1)  \beta}
a_{({\bf p}_1 - \frac1{2}{\bf q}_1 +  \frac1{2}{\bf q})\gamma}^\dag
a_{({\bf k}_1 - \frac1{2} {\bf q}_1) \delta} .
\end{equation}
 Matrix elements of $\alpha=\delta$ and $\beta=\gamma$ are
relevant for CBO channels, and they are the same as those given by
\begin{eqnarray}\label{EqEl-Ph-CBO2}
{\cal P}_\Gamma^\prime ({\bf q}) &=& 
\frac1{2N}\sum_{{\bf k}_1{\bf p}_1{\bf q}_1} 
\eta_{\Gamma}({\bf k}_1-{\bf p}_1) 
\frac1{8} \left[
- 3 \rho_{c}({\bf k}_1+
\mbox{$\frac1{2}$}{\bf q}_1+\mbox{$\frac1{2}$}{\bf q},
{\bf k}_1 - \mbox{$\frac1{2}$}{\bf q}_1)
\rho_{c}({\bf p}_1 - 
\mbox{$\frac1{2}$}{\bf q}_1+\mbox{$\frac1{2}$}{\bf q},
{\bf p}_1 + \mbox{$\frac1{2}$}{\bf q}_1) \right.
\nonumber \\ && \hspace*{2cm} \left. 
+ \rho_{s}({\bf k}_1+
\mbox{$\frac1{2}$}{\bf q}_1+\mbox{$\frac1{2}$}{\bf q},
{\bf k}_1 - \mbox{$\frac1{2}$}{\bf q}_1)
\rho_{s}({\bf p}_1 - 
\mbox{$\frac1{2}$}{\bf q}_1+\mbox{$\frac1{2}$}{\bf q},
{\bf p}_1 + \mbox{$\frac1{2}$}{\bf q}_1)
\right]  + \cdots ,
\end{eqnarray}
with
\begin{equation}
\rho_{c}({\bf k}_1,{\bf k}_2)
= a_{{\bf k}_1\uparrow}^{\dag}a_{{\bf k}_2\uparrow}
+ a_{{\bf k}_1\downarrow}^{\dag}a_{{\bf k}_2\downarrow},
\qquad
\rho_{s}({\bf k}_1,{\bf k}_2)
= a_{{\bf k}_1\uparrow}^{\dag}a_{{\bf k}_2\uparrow}
- a_{{\bf k}_1\downarrow}^{\dag}a_{{\bf k}_2\downarrow} ,
\end{equation}
except for those given by what appear through the commutation of
operators, which are not shown here.
This expression is useful to obtain Eqs.~(\ref{EqS-CBO}) and (\ref{EqZCBO}).
When we  replace variables in Eq.~(\ref{EqEl-Ph-CBO}) in such a way that
${\bf k}_1 + \frac1{2} {\bf q}_1 = {\bf k}_2 + \frac1{2}{\bf q}_2$, 
${\bf p}_1 + \frac1{2} {\bf q}_1 = {\bf p}_2 + \frac1{2} {\bf q}_2$,  
and
${\bf k}_1 - \frac1{2} {\bf q}_1 = - {\bf p}_2 +\frac1{2}{\bf q}_2$, 
Eq.~(\ref{EqEl-Ph-CBO}) turns out to
\begin{equation}\label{EqEl-Ph-SC}
{\cal P}_\Gamma ({\bf q}) =
\frac1{2N}\sum_{{\bf k}_2{\bf p}_2{\bf q}_2} 
\sum_{\alpha\beta\gamma\delta}
\eta_{\Gamma}({\bf k}_2-{\bf p}_2) 
\left({\bf s}^{\alpha\beta} \!\cdot {\bf s}^{\gamma\delta}\right)
a_{({\bf k}_2 +  \frac1{2}{\bf q}_2 +  \frac1{2}{\bf q})\alpha}^\dag 
a_{({\bf p}_2 + \frac1{2}{\bf q}_2)  \beta}
a_{(-{\bf k}_2 + \frac1{2}{\bf q}_2 +  \frac1{2}{\bf q})\gamma}^\dag
a_{(-{\bf p}_2 + \frac1{2} {\bf q}_2) \delta} . 
\end{equation}
Matrix elements of $\gamma=-\alpha$ and $\delta=-\beta$ are
relevant for singlet SC channels, and they are the same as those given by
\begin{eqnarray}\label{EqEl-Ph-SC2}
{\cal P}_\Gamma^{\prime\prime} ({\bf q}) &=&
\frac1{2N}\sum_{{\bf k}_2{\bf p}_2{\bf q}_2} 
\eta_{\Gamma}({\bf k}_2-{\bf p}_2) 
\frac1{8} \left[
- 3
\rho_{1}^{\dag}({\bf k}_2 + \mbox{$\frac1{2}$}{\bf q}_2 
+  \mbox{$\frac1{2}$} {\bf q},
-{\bf k}_2 + \mbox{$\frac1{2}$}{\bf q}_2 
+  \mbox{$\frac1{2}$} {\bf q})
\rho_{1}({\bf p}_2 + \mbox{$\frac1{2}$}{\bf q}_2,
{\bf p}_2 + \mbox{$\frac1{2}$}{\bf q}_2 )
\right.
\nonumber \\ && \hspace*{1cm} \left.
+  
\rho_{3}^{\dag}({\bf k}_2 + \mbox{$\frac1{2}$}{\bf q}_2 
+  \mbox{$\frac1{2}$} {\bf q},
-{\bf k}_2 + \mbox{$\frac1{2}$}{\bf q}_2 
+  \mbox{$\frac1{2}$} {\bf q})
\rho_{3}({\bf p}_2 + \mbox{$\frac1{2}$}{\bf q}_2,
- {\bf p}_2 + \mbox{$\frac1{2}$}{\bf q}_2)\right]  
 + \cdots,
\end{eqnarray}
with
\begin{equation}
\rho_{1}^\dag({\bf k}_1,{\bf k}_2) = 
a_{{\bf k}_1\uparrow}^\dag a_{{\bf k}_2\downarrow}^\dag
-a_{{\bf k}_1\downarrow}^\dag a_{{\bf k}_2\uparrow}^\dag ,
\qquad
\rho_{3}^\dag({\bf k}_1,{\bf k}_2) = 
a_{{\bf k}_1\uparrow}^\dag a_{{\bf k}_2\downarrow}^\dag
+a_{{\bf k}_1\downarrow}^\dag a_{{\bf k}_2\uparrow}^\dag ,
\end{equation}\end{widetext} 
except for those given by what appear through the commutation of
operators, which are not shown here. 
This expression is useful to obtain Eqs.~(\ref{EqS-SC}) and (\ref{EqZSC}).
The following relation is also useful:
\begin{eqnarray}
2\eta_{s}({\bf k}-{\bf p}) &=& 
\eta_s({\bf k})\eta_s({\bf p})
+ \eta_d({\bf k})\eta_d({\bf p})
\nonumber \\ && \quad
+ \eta_x({\bf k})\eta_x({\bf p})
+ \eta_y({\bf k})\eta_y({\bf p}) . \qquad
\end{eqnarray}
The factor $\eta_{s}({\bf k}-{\bf p})$ appearing in
Eqs.~(\ref{EqEl-Ph-CBO}), (\ref{EqEl-Ph-CBO2}),
(\ref{EqEl-Ph-SC})  and (\ref{EqEl-Ph-SC2}) can be
decoupled in this way. Even if only the part of
$\Gamma=s$ in Eqs.~(\ref{EqElPhP}) and (\ref{EqElPhD}) is considered,
there appear contributions from $d\gamma$ and $p$ waves.


\begin{thebibliography}{}
\bibitem{ComSlave}
When we take the other statistics different from those of the slave-boson
$t$-$J$ model, fermionic $e$ and bosonic $c$ particles, results in the MF
approximation for such a model are totally different from those for the
slave-boson $t$-$J$ model. One of the most remarkable differences is that the
volumes of the Fermi surface are totally different between the two models. 
This is another evidence that the MF approximation for the auxiliary-particle
models, including the slave boson
$t$-$J$ model, is irrelevant to describe dynamics of electrons in the
original $t$-$J$ model.
\end{thebibliography}
\end{document}